\documentclass[conference]{IEEEtran}
\IEEEoverridecommandlockouts
\usepackage{amsmath,amssymb,amsfonts}
\usepackage{algorithmic}
\usepackage{algorithm}
\usepackage{subcaption}
\usepackage{array}
\usepackage{textcomp}
\usepackage{stfloats}
\usepackage{url}
\usepackage{multirow}
\usepackage{verbatim}
\usepackage{enumitem}
\usepackage{xcolor}
\usepackage{graphicx}
\usepackage{cite}
\def\BibTeX{{\rm B\kern-.05em{\sc i\kern-.025em b}\kern-.08em
    T\kern-.1667em\lower.7ex\hbox{E}\kern-.125emX}}
\begin{document}

\title{Stateful Logic In-Memory Using Gain-Cell eDRAM\\
\thanks{This work was supported by the European Research Council through the European Union's Horizon 2020 Research and Innovation Programme under Grant 757259 and by the European Research Council through the European Union's Horizon Europe Research and Innovation Programme under Grant 101069336.}
}

\author{\IEEEauthorblockN{Barak Hoffer and Shahar Kvatinsky}
\IEEEauthorblockA{\textit{Andrew and Erna Viterbi Faculty of Electrical and Computer Engineering} \\
\textit{Technion - Israel Institute of Technology}\\
Haifa, Israel \\
barakhoffer@campus.technion.ac.il, shahar@ee.technion.ac.il}
}

\maketitle
\begin{abstract}
Modern data-intensive applications demand memory solutions that deliver high-density, low-power, and integrated computational capabilities to reduce data movement overhead. This paper presents the use of Gain-Cell embedded DRAM (GC-eDRAM)---a compelling alternative to traditional SRAM and eDRAM---for stateful, in-memory logic. We propose a circuit design that exploits GC-eDRAM’s dual-port architecture and nondestructive read operation to perform logic functions directly within the GC-eDRAM memory array. Our simulation results demonstrate a 5us retention time coupled with a 99.5\% success rate for computing the logic gates. By incorporating processing-in-memory (PIM) functionality into GC-eDRAM, our approach enhances memory and compute densities, lowers power consumption, and improves overall performance for data-intensive applications.
\end{abstract}

\begin{IEEEkeywords}
stateful-logic, processing-in-memory, eDRAM, GC-eDRAM
\end{IEEEkeywords}

\section{Introduction}
\IEEEPARstart{A}{s} 
modern applications grow increasingly complex and data-intensive, embedded memories occupy a growing fraction of the silicon area in system-on-chips (SoCs). Six-transistor (6T) static random-access memory (SRAM) technology is often the preferred choice for embedded memory due to its robustness and speed. 
Yet, its large size and high static power consumption often dominate both the area and power budgets of SoCs \cite{Hennessy2017}. Although traditional 1T-1C embedded DRAM (eDRAM) offers a higher-density alternative, the additional process steps required to fabricate the capacitor, coupled with reliability concerns in deeply-scaled process nodes, have limited its widespread use.
Gain-cell (GC) eDRAM has emerged as a promising alternative, offering higher density and lower leakage power consumption than SRAM, while retaining compatibility with standard CMOS processes\cite{Meinerzhagen2018,Bonetti2020,Shalom2018,Harel2022,Singh2023,Golman2024}. By decoupling the read and write paths, GC-eDRAM inherently supports dual-port operation, which enables faster access times compared to 1T-1C DRAM. Additionally, nondestructive reads eliminate the need for frequent data rewrites, further reducing operational overhead.

Traditionally, scaling up on-chip memory has been the principal strategy for boosting the performance of data-centric applications. However, the associated increase in data movement between memory and processors introduces high energy consumption, latency and bandwidth constraints, ultimately limiting overall system performance. Processing-in-memory (PIM) has emerged as a powerful approach to overcome these bottlenecks by moving computational tasks directly into the memory fabric, thereby decreasing data transfers, reducing energy costs, and relieving pressure on the memory hierarchy~\cite{Dong2017,Eckert2018,Seshadri2017}. PIM has been extensively studied for operations such as Multiply-Accumulate (MAC) operations~\cite{Khwa2024a}, Boolean logic~\cite{Gao2019}, and content-addressable memory (CAM) operations~\cite{Jahshan2023}.

In this paper, we extend the functionality of GC-eDRAM by introducing PIM functionality through stateful logic, a concept  previously demonstrated in memristors~\cite{Borghetti2010,Reuben2017,BenHur2020,Leitersdorf2022,Perach2024}. Stateful logic enables in-memory execution of Boolean logic operations, facilitating parallelism and local computations---both of which significantly speed up tasks and improve overall system throughput. By adopting GC-eDRAM with our proposed method in place of traditional SRAM or DRAM for data-intensive applications, designers can achieve increased memory and compute densities, reduced power consumption, and effective PIM capabilities. We specifically propose a 3T NMOS GC array capable of executing stateful NOR and NOT logic gates in memory. Using Monte Carlo simulations, we verify the robustness of these operations in the presence of process variation and device mismatch, achieving a 99.5\% success rate. 

\begin{figure}[bt]
  \centering
  \begin{subfigure}{.46\columnwidth}
  \centering
  \includegraphics[width=\columnwidth]{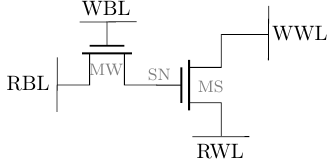}
  \caption{}
  \end{subfigure}
  \hspace{8pt}
  \begin{subfigure}{.48\columnwidth}
  \centering
  \includegraphics[width=\columnwidth]{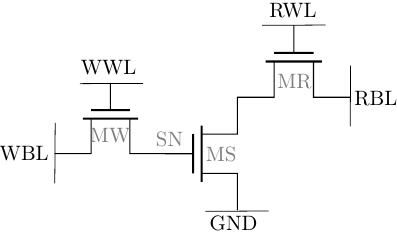}
  \caption{}
  \end{subfigure}
  \hfill
  \begin{subfigure}{.48\columnwidth}
  \centering
  \includegraphics[width=\columnwidth]{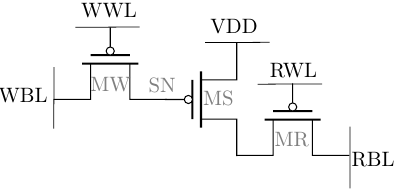}
  \caption{}
  \end{subfigure}
  \caption{Different types of GC-eDRAM memory bitcells: \\ (a) 2T NMOS GC, (b) 3T NMOS GC, (c) 3T PMOS GC}
  \label{fig:gc-structure}
  \end{figure}

\begin{figure*}[ht]
\centering
  \begin{subfigure}{0.31\textwidth}
      \centering
      \includegraphics{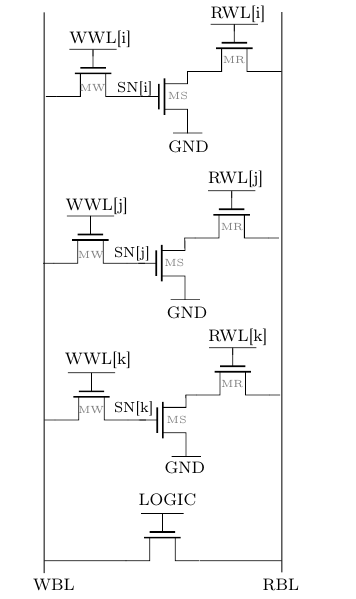}
      \caption{}
  \end{subfigure} 
  \begin{subfigure}{0.31\textwidth}
      \centering
      \includegraphics{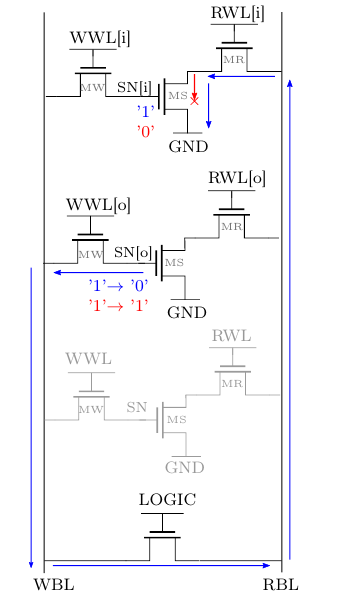}
      \caption{}
  \end{subfigure}
  \begin{subfigure}{0.31\textwidth}
      \centering
      \includegraphics{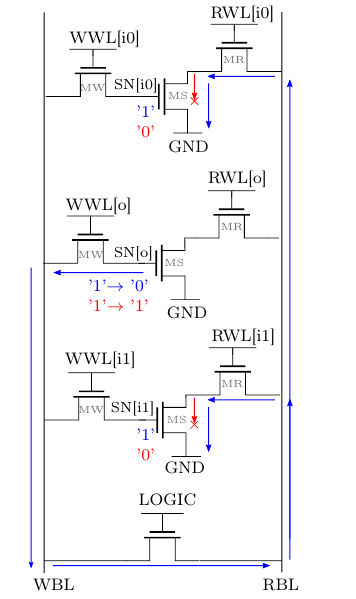}
      \caption{}
  \end{subfigure}

  \caption{GC-eDRAM stateful logic in-memory implementation: (a) RBL to WBL connection using a transistor, (b) Stateful in-memory NOT logic operation, (c) Stateful in-memory NOR logic operation.}
  \label{fig:logic}
  \end{figure*}

\section{Embedded DRAM Gain-Cell Memory}
Each memory cell in gain-cell eDRAM is composed of multiple transistors. The transistor capacitance replaces the capacitor used in the 1T-1C cells in traditional DRAM. Sense amplifiers are employed to amplify the stored signal during read operations, while additional peripheral circuits manage key functions such as address decoding, timing control, and power management.
Several topologies for GC-eDRAM bitcells have been proposed, ranging from two to four transistor circuits \cite{Shalom2018}.
Each topology generally includes three main components:

\begin{enumerate}
    \item A \textit{write port} utilizing a write word line (WWL) to enable data input, driving the write bit line (WBL) into the memory cell.
    \item A \textit{storage node} (SN), storing data through  the parasitic capacitance of surrounding transistors and interconnect.
    \item A \textit{read port} that amplifies the voltage stored on the SN and drives it as current through the read bit line (RBL) when the read word line (RWL) is activated. 
\end{enumerate}
The simplest topology is the 2T GC (Fig. \ref{fig:gc-structure}a), consisting of a write transistor (MW) and a read/storage  transistor (MS).
Adding another read transistor (MR) and turning the storage transistor (MS) to a dedicated device, the 3T GC (Fig. \ref{fig:gc-structure}b-c) improves robustness, speed, and retention time. However, this enhancement comes at the cost of increased area and the need for each bitcell to connect to either GND in the NMOS version or $V_{DD}$ in the PMOS version.
 
\section{Stateful Logic for GC-eDRAM}
\subsection{Proposed Technique}
To enable stateful logic in GC-eDRAM, we introduce a single transistor for each bitline of the array, allowing optional connection of the RBL to the WBL, as depicted in Fig. \ref{fig:logic}a. 

The logic operation is performed in-memory using a two-step write pulse process:
\begin{enumerate}
    \item The WBL, LOGIC line, and WWL of the output are raised to $V_{DD}$, charging the SN of the output bitcell and initializing it to logical '1'.
    \item While the WWL and LOGIC lines remain high, the WBL driver is gated, and the input is selected by driving one or more RWL of input rows.
\end{enumerate}
The final state of the output SN is determined by the states of the input bitcells, effectively computing the result of the logic operation.

This technique constructs concurrent n-bit logic operations across all columns, using inputs from the selected rows to generate  an output in the target row. More complex logic functions can be realized by performing sequential logic operations, where the output bitcell of one gate serves as the input to the next. Such bulk bit-wise processing can accelerate various computational workloads, as shown in earlier studies~\cite{Leitersdorf2022,Perach2024}. In the following subsections, we detail the design of two fundamental logic gates: NOT and NOR. Together, they form a functionally complete set, allowing synthesis tools to determine the required sequence of operations for any desired logic function~\cite{BenHur2020}. Although our examples use a 3T NMOS GC topology, the same approach can be applied to other GC-eDRAM bitcell configurations (such as 2T and 3T PMOS), potentially yielding different logic gates (e.g., NAND, OR, AND) depending on the structure.

\begin{figure}[t]
    \centering
    \includegraphics[width=0.75\columnwidth]{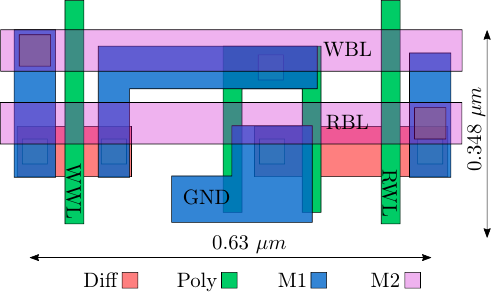}
    \caption{3T NMOS GC-eDRAM bitcell layout in GF 28nm technology.}
    \label{fig:layout}
    \vspace{-13pt}
  \end{figure}

\subsection{NOT gate}
The NOT logic operation inverts a single input, as shown in Fig. \ref{fig:logic}b. The operation begins by initializing to logical '1'. If the input stores a '1', MR becomes active, and when the RWL of the input is driven high, a discharge current flows through the WBL and the MW of the output, discharging its SN. If the input stores a '0', MR remains inactive, preventing any discharge. In both cases, the resulting state stored in the output SN is the inversion of the input state, successfully completing the NOT operation.

\subsection{NOR gate}
The NOR gate is implemented by extending the NOT gate to include an additional input, and it outputs '1' only when all inputs are '0'.
The NOR gate in the 3T GC-eDRAM structure is shown in Fig. \ref{fig:logic}c. To perform the NOR operation, the output is first initialized to '1', and the same voltage signals used in the NOT gate are applied, with an additional input row also active. If any of the input SNs store '1', a discharge current will flow, causing the output SN to switch to '0'. If all input SNs store '0', no discharge occurs, and the output remains at '1', completing the NOR operation.

\begin{figure}[t]
    \centering
    \includegraphics[width=.75\columnwidth]{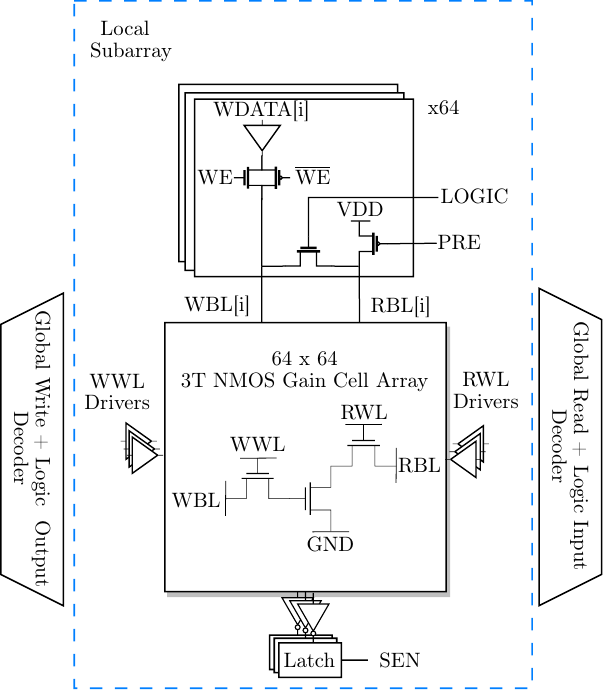}

  \caption{Hierarchial GC-eDRAM memory architecture with in-memory stateful logic support.}
  \label{fig:arch}
    \vspace{-12pt}
\end{figure}

\section{Architecture and Implementation}
\subsection{3T NMOS gain-cell array}
A 3T NMOS GC-eDRAM cell was implemented in GlobalFoundries (GF) 28nm technology. It employs multi-threshold voltage ($V_t$) devices for optimized performance. The WWL selector (MW) utilizes a high-threshold voltage (HVT) transistor to minimize leakage current, while the read port (MR) and storage node (MS) use low-threshold voltage (LVT) transistors to reduce access latency. The bitcell is measured at 0.63µm $\times$ 0.348µm, featuring WWL and RWL lines routed horizontally in polysilicon, and the RBL and WBL lines routed in M2, as depicted in Fig. \ref{fig:layout}.
\vspace{-0.1cm}
\subsection{Memory Architecture}
A hierarchical eDRAM array architecture~\cite{Bonetti2020}, is employed, consisting of 64$\times$64 sub-arrays, as depicted in Fig. \ref{fig:arch}. Each sub-array has its own dedicated set of WWL/RWL drivers, while the row selection for read/write/logic is managed by global decoders. The use of sub-arrays with a relatively low number of rows simplifies operation and enhances the performance of logic operations by maintaining a short discharge path. This hierarchical memory structure also promotes greater parallelism, as each sub-array can execute logic operations in its own row, with all sub-arrays operating in parallel and sharing common control logic.

Each column in the sub-array includes a WBL driver, a pass gate, a precharge PMOS for the read operations, and an NMOS to connect the RBL and WBL during logic operations. The area overhead for the additional NMOS per bitline is 0.5\%. For data reading, an inverter and latch-based sense amplifier (SA) is connected to the RBL of each column. The timing for the read pulse is set at 3ns, and the write pulse is set at 1ns. The time-after-write is the time passed since the write operation. The data retention time (DRT) is defined as 15{\textmu}s using these read/write pulse timings, by finding the maximum time-after-write, where in the following read operation, the read value remains correct. The logic pulse time is set as 3ns, with the first 1ns used to drive the output SN to '1' via the WBL, followed by 2ns, where the RWL of the inputs is asserted high to evaluate the logic function.
\vspace{-0.2cm}
\section{Results}
\vspace{-0.1cm}
\begin{figure}[t]
  \centering
  \includegraphics[width=.75\columnwidth]{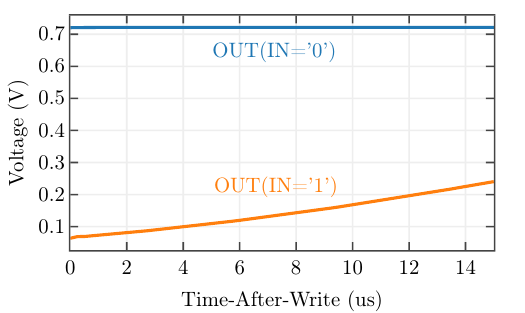}
  \caption{Voltage across the output SN as a function of time-after-write of the input, for the NOT logic gate.}
  \label{fig:logic_drt}
  \vspace{-8pt}
\end{figure}

\begin{figure}[t]
  \centering
  \includegraphics[width=.75\columnwidth]{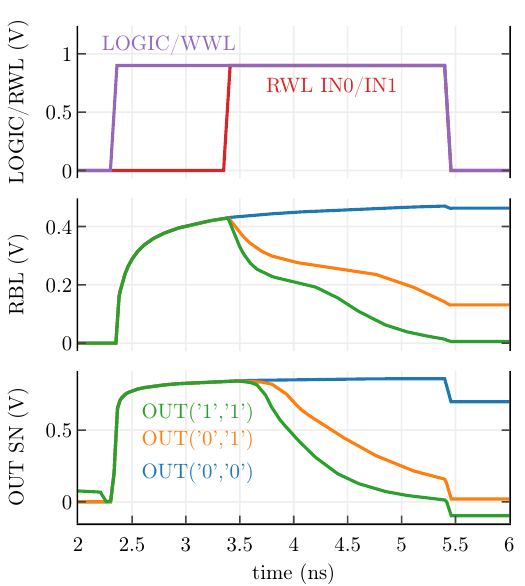}
  \caption{Waveforms extracted from simulation, demonstrating the stateful NOR logic operation using a two-step write pulse process.}
  \label{fig:waves}
  \vspace{-15pt}
\end{figure}

To evaluate and demonstrate stateful logic operations for GC-eDRAM, we designed the memory sub-array in GF 28nm using Cadence Virtuoso and simulated the NOR and NOT operations within the array. 

The NOT logic operation was simulated for both input cases, while varying the time elapsed after writing the input, as depicted in Fig. \ref{fig:logic_drt}. The results indicate correct logic output based on the output SN voltage corresponding to each input state. The voltage representing a logical '1' increases as a function of time-after-write for the input. However, the operation does not achieve full voltage swing. When the input is '1' and the output SN discharges after initialization, the final voltage depends on the discharge of the input SN, which weakens the driving strength of the input MS transistor over time. To ensure high success rate for the logic operation, the DRT is reduced to 5{\textmu}s, which is comparable to other works that use similar technology nodes~\cite{Yigit2023, He2024b}. With this reduced DRT, refreshing an entire 64-row sub-array takes 256ns, maintaining memory availability at a relatively high rate of 95\%. The DRT can be further improved by implementing the GC memory with mixed-type transistors~\cite{Yigit2023, Narinx2019}.

The NOR operation waveforms, extracted from simulations, are depicted in Fig. \ref{fig:waves}. The figure shows the two stages of the logic pulse: first, both WWL and LOGIC are raised, allowing the RBL and WBL to charge. Then, the RWL of the inputs is activated while WWL and LOGIC remain high. The RBL initially charges, then discharges or remains high depending on the inputs. The output SN voltage correctly reflects the logic result, staying high only when both inputs are '0', as required from a NOR operation.

The energy consumption of the proposed stateful in-memory NOT and NOR operations was measured through simulation, yielding 13.4~fJ and 13.5~fJ, respectively. For reference, the energy consumption for standard read and write operations was measured at 13.3~fJ and 5.7~fJ, respectively. In a traditional flow without PIM, executing a logic function necessitates separate read, compute, and write-back steps, incurring the combined energy of all three. By contrast, the stateful logic approach integrates the write-back step into the logic operation itself, eliminating any additional energy cost. Furthermore, sensing two cells simultaneously consumes 13.34~fJ, providing a benchmark for sensing-based logic schemes that interpret the SA's output as the logic result. While sensing-based logic requires a write-back operation, as in traditional flow without PIM.

To assess the robustness of the logic operations under process and device variations, we conducted 1K Monte Carlo simulations for each input combination of both gates, at room temperature with $V_{DD}=0.9V$. As shown in Tables \ref{tab:not_mc},\ref{tab:nor_mc}, the success rate was consistently high, with 99.5\% success rate for the worst-case scenario, where only one input stores logical '1'. In these cases, the failures are caused by a faster input SN discharge combined with a higher threshold voltage of the SA.

\begin{table}[t]
  \caption{NOT Monte Carlo Results}
  \vspace{-10pt}
  \label{tab:not-success}
  \begin{center}
    \normalsize
    \def\arraystretch{1.03}
    \begin{tabular}{|c|c|c|c|}
      \hline
      \textbf{IN} & \textbf{OUT} & \textbf{Success Rate} \\ \hline \hline
      `0'          & `1'          & 100\%       \\ \hline
      `1'          & `0'          & 99.5\%        \\ \hline
    \end{tabular}
  \end{center}
  \vspace{-7pt}
  \label{tab:not_mc}
\end{table}

\begin{table}[t]
  \caption{NOR Monte Carlo Results}
  \vspace{-10pt}
  \label{tab:nor-success}
  \begin{center}
    \normalsize
    \def\arraystretch{1.03}
    \begin{tabular}{|c|c|c|c|}
      \hline
      \textbf{IN1} & \textbf{IN0} & \textbf{OUT} & \textbf{Success Rate} \\ \hline \hline
      `0'          & `0'          & `1'          & 100\%       \\ \hline
      `0'          & `1'          & `0'          & 99.5\%        \\ \hline
      `1'          & `0'          & `0'          & 99.5\%        \\ \hline
      `1'          & `1'          & `0'          & 100\%        \\ \hline
    \end{tabular}
  \end{center}
  \vspace{-20pt}
  \label{tab:nor_mc}
\end{table}
\section{Conclusion}
In this paper, we presented a novel in-memory stateful logic approach using GC-eDRAM. We began by describing the necessary modifications to the array structure that enable stateful logic, followed by the design and implementation of  a 3T NMOS-based GC array to validate our method. Our results demonstrated successful in-memory NOT and NOR gate operations, emphasizing the importance of limiting the number of rows in the array and reduceing effective data retention times to ensure high success rates. Moving forward, we plan to extend this work by exploring additional computational workloads that can benefit from GC-eDRAM stateful logic, further leveraging its potential for high-density and energy-efficient processing.

\bibliographystyle{IEEEtran}
\bibliography{IEEEabrv,ref.bib}

% Generated by IEEEtran.bst, version: 1.14 (2015/08/26)
\begin{thebibliography}{10}
\providecommand{\url}[1]{#1}
\csname url@samestyle\endcsname
\providecommand{\newblock}{\relax}
\providecommand{\bibinfo}[2]{#2}
\providecommand{\BIBentrySTDinterwordspacing}{\spaceskip=0pt\relax}
\providecommand{\BIBentryALTinterwordstretchfactor}{4}
\providecommand{\BIBentryALTinterwordspacing}{\spaceskip=\fontdimen2\font plus
\BIBentryALTinterwordstretchfactor\fontdimen3\font minus \fontdimen4\font\relax}
\providecommand{\BIBforeignlanguage}[2]{{%
\expandafter\ifx\csname l@#1\endcsname\relax
\typeout{** WARNING: IEEEtran.bst: No hyphenation pattern has been}%
\typeout{** loaded for the language `#1'. Using the pattern for}%
\typeout{** the default language instead.}%
\else
\language=\csname l@#1\endcsname
\fi
#2}}
\providecommand{\BIBdecl}{\relax}
\BIBdecl

\bibitem{Hennessy2017}
J.~L. Hennessy and D.~A. Patterson, \emph{Computer Architecture: a Quantitative Approach}, 6th~ed.\hskip 1em plus 0.5em minus 0.4em\relax Morgan Kaufmann, 2017.

\bibitem{Meinerzhagen2018}
P.~Meinerzhagen, A.~Teman, R.~Giterman, N.~Edri, A.~Burg, and A.~Fish, \emph{Gain-cell Embedded DRAMs for Low-power VLSI Systems-on-chip}.\hskip 1em plus 0.5em minus 0.4em\relax Springer, 2018.

\bibitem{Bonetti2020}
A.~Bonetti, R.~Golman, R.~Giterman, A.~Teman, and A.~Burg, ``Gain-{{Cell Embedded DRAMs}}: {{Modeling}} and {{Design Space}},'' \emph{IEEE Transactions on Very Large Scale Integration (VLSI) Systems}, vol.~28, no.~3, pp. 646--659, Mar. 2020.

\bibitem{Shalom2018}
A.~Shalom, R.~Giterman, and A.~Teman, ``High {{Density GC-eDRAM Design}} in 16nm {{FinFET}},'' in \emph{2018 25th {{IEEE International Conference}} on {{Electronics}}, {{Circuits}} and {{Systems}} ({{ICECS}})}, Dec. 2018, pp. 585--588.

\bibitem{Harel2022}
O.~Harel, E.~N. Casarrubias, M.~Eggimann, F.~G{\"u}rkaynak, L.~Benini, A.~Teman, R.~Giterman, and A.~Burg, ``64-{{kB}} 65-nm {{GC-eDRAM With Half-Select Support}} and {{Parallel Refresh Technique}},'' \emph{IEEE Solid-State Circuits Letters}, vol.~5, pp. 170--173, 2022.

\bibitem{Singh2023}
S.~Singh, N.~Surana, K.~Prasad, P.~Jain, J.~Mekie, and M.~Awasthi, ``{{HyGain}}: {{High-performance}}, {{Energy-efficient Hybrid Gain Cell-based Cache Hierarchy}},'' \emph{ACM Transactions on Architecture and Code Optimization}, vol.~20, no.~2, pp. 24:1--24:20, Mar. 2023.

\bibitem{Golman2024}
R.~Golman, A.~Segev, and A.~Teman, ``A {{4T GC-eDRAM Bitcell}} with {{Differential Readout Mechanism For High Performance Applications}},'' in \emph{2024 19th {{Conference}} on {{Ph}}.{{D Research}} in {{Microelectronics}} and {{Electronics}} ({{PRIME}})}, Jun. 2024, pp. 1--4.

\bibitem{Dong2017}
Q.~Dong, S.~Jeloka, M.~Saligane, Y.~Kim, M.~Kawaminami, A.~Harada, S.~Miyoshi, D.~Blaauw, and D.~Sylvester, ``A 0.{{3V VDDmin}} 4+{{2T SRAM}} for searching and in-memory computing using 55nm {{DDC}} technology,'' in \emph{2017 {{Symposium}} on {{VLSI Circuits}}}, Jun. 2017, pp. C160--C161.

\bibitem{Eckert2018}
C.~Eckert, X.~Wang, J.~Wang, A.~Subramaniyan, R.~Iyer, D.~Sylvester, D.~Blaauw, and R.~Das, ``Neural cache: Bit-serial in-cache acceleration of deep neural networks,'' in \emph{Proceedings of the 45th {{Annual International Symposium}} on {{Computer Architecture}}}, ser. {{ISCA}} '18.\hskip 1em plus 0.5em minus 0.4em\relax Los Angeles, California: IEEE Press, Jun. 2018, pp. 383--396.

\bibitem{Seshadri2017}
V.~Seshadri, D.~Lee, T.~Mullins, H.~Hassan, A.~Boroumand, J.~Kim, M.~A. Kozuch, O.~Mutlu, P.~B. Gibbons, and T.~C. Mowry, ``Ambit: In-memory accelerator for bulk bitwise operations using commodity {{DRAM}} technology,'' in \emph{Proceedings of the 50th {{Annual IEEE}}/{{ACM International Symposium}} on {{Microarchitecture}}}, ser. {{MICRO-50}} '17.\hskip 1em plus 0.5em minus 0.4em\relax New York, NY, USA: Association for Computing Machinery, Oct. 2017, pp. 273--287.

\bibitem{Khwa2024a}
W.-S. Khwa, P.-C. Wu, J.-J. Wu, J.-W. Su, H.-Y. Chen, Z.-E. Ke, T.-C. Chiu, J.-M. Hsu, C.-Y. Cheng, Y.-C. Chen, C.-C. Lo, R.-S. Liu, C.-C. Hsieh, K.-T. Tang, and M.-F. Chang, ``34.2 {{A}} 16nm {{96Kb Integer}}/{{Floating-Point Dual-Mode-Gain-Cell-Computing-in-Memory Macro Achieving}} 73.3-163.{{3TOPS}}/{{W}} and 33.2-91.{{2TFLOPS}}/{{W}} for {{AI-Edge Devices}},'' in \emph{2024 {{IEEE International Solid-State Circuits Conference}} ({{ISSCC}})}, vol.~67, Feb. 2024, pp. 568--570.

\bibitem{Gao2019}
F.~Gao, G.~Tziantzioulis, and D.~Wentzlaff, ``{{ComputeDRAM}}: {{In-Memory Compute Using Off-the-Shelf DRAMs}},'' in \emph{Proceedings of the 52nd {{Annual IEEE}}/{{ACM International Symposium}} on {{Microarchitecture}}}, ser. {{MICRO}} '52.\hskip 1em plus 0.5em minus 0.4em\relax New York, NY, USA: Association for Computing Machinery, Oct. 2019, pp. 100--113.

\bibitem{Jahshan2023}
Z.~Jahshan, I.~Merlin, E.~Garz{\'o}n, and L.~Yavits, ``{{DASH-CAM}}: {{Dynamic Approximate SearcH Content Addressable Memory}} for genome classification,'' in \emph{Proceedings of the 56th {{Annual IEEE}}/{{ACM International Symposium}} on {{Microarchitecture}}}, ser. {{MICRO}} '23.\hskip 1em plus 0.5em minus 0.4em\relax New York, NY, USA: Association for Computing Machinery, Dec. 2023, pp. 1453--1465.

\bibitem{Borghetti2010}
J.~Borghetti, G.~S. Snider, P.~J. Kuekes, J.~J. Yang, D.~R. Stewart, and R.~S. Williams, ```{{Memristive}}' switches enable `stateful' logic operations via material implication,'' \emph{Nature}, vol. 464, no. 7290, pp. 873--876, Apr. 2010.

\bibitem{Reuben2017}
J.~Reuben, R.~{Ben-Hur}, N.~Wald, N.~Talati, A.~H. Ali, P.-E. Gaillardon, and S.~Kvatinsky, ``Memristive logic: {{A}} framework for evaluation and comparison,'' in \emph{2017 27th {{International Symposium}} on {{Power}} and {{Timing Modeling}}, {{Optimization}} and {{Simulation}} ({{PATMOS}})}, Sep. 2017, pp. 1--8.

\bibitem{BenHur2020}
R.~Ben-Hur, R.~Ronen, A.~Haj-Ali, D.~Bhattacharjee, A.~Eliahu, N.~Peled, and S.~Kvatinsky, ``Simpler magic: Synthesis and mapping of in-memory logic executed in a single row to improve throughput,'' \emph{IEEE Transactions on Computer-Aided Design of Integrated Circuits and Systems}, vol.~39, no.~10, pp. 2434--2447, 2020.

\bibitem{Leitersdorf2022}
O.~Leitersdorf, R.~Ronen, and S.~Kvatinsky, ``{{MatPIM}}: {{Accelerating Matrix Operations}} with {{Memristive Stateful Logic}},'' in \emph{2022 {{IEEE International Symposium}} on {{Circuits}} and {{Systems}} ({{ISCAS}})}, May 2022, pp. 215--219.

\bibitem{Perach2024}
B.~Perach, R.~Ronen, B.~Kimelfeld, and S.~Kvatinsky, ``Understanding {{Bulk-Bitwise Processing In-Memory Through Database Analytics}},'' \emph{IEEE Transactions on Emerging Topics in Computing}, vol.~12, no.~1, pp. 7--22, Jan. 2024.

\bibitem{Yigit2023}
A.~Yigit, E.~N. Casarrubias, R.~Giterman, and A.~Burg, ``A 128-kbit {{GC-eDRAM With Negative Boosted Bootstrap Driver}} for 11.3{\texttimes} {{Lower-Refresh Frequency}} at a 2.5\% {{Area Overhead}} in 28-nm {{FD-SOI}},'' \emph{IEEE Solid-State Circuits Letters}, vol.~6, pp. 13--16, 2023.

\bibitem{He2024b}
Y.~He, S.~Fan, X.~Li, L.~Lei, W.~Jia, C.~Tang, Y.~Li, Z.~Huang, Z.~Du, J.~Yue, X.~Li, H.~Yang, H.~Jia, and Y.~Liu, ``34.7 {{A}} 28nm 2.{{4Mb}}/mm2 6.9 - 16.{{3TOPS}}/mm2 {{eDRAM-LUT-Based Digital-Computing-in-Memory Macro}} with {{In-Memory Encoding}} and {{Refreshing}},'' in \emph{2024 {{IEEE International Solid-State Circuits Conference}} ({{ISSCC}})}, vol.~67, Feb. 2024, pp. 578--580.

\bibitem{Narinx2019}
J.~Narinx, R.~Giterman, A.~Bonetti, N.~Frigerio, C.~Aprile, A.~Burg, and Y.~Leblebici, ``A 24 kb {{Single-Well Mixed 3T Gain-Cell eDRAM}} with {{Body-Bias}} in 28 nm {{FD-SOI}} for {{Refresh-Free DSP Applications}},'' in \emph{2019 {{IEEE Asian Solid-State Circuits Conference}} ({{A-SSCC}})}, Nov. 2019, pp. 219--222.

\end{thebibliography}
\end{document}